\documentclass[]{style/iromlab}

\DeclareDocumentEnvironment{example}{}{\noindent\textbf{Running example:}\itshape}{}

\usepackage{ifthen}
\newboolean{include-notes}
\newboolean{include-new}
\newboolean{include-remove}
\setboolean{include-notes}{true}
\setboolean{include-new}{false}
\setboolean{include-remove}{false}

\usepackage[dvipsnames]{xcolor}
\usepackage[normalem]{ulem}
\newcommand{\justin}[1]{\ifthenelse{\boolean{include-notes}}{\textcolor{orange}{\textbf{Jaime:} #1}}{}}

\newcommand{\princeton}[1]{\ifthenelse{\boolean{include-notes}}{\textcolor{orange}{#1}}{}}

\usepackage{multicol}
\usepackage{amsmath, amsfonts, amssymb, amsthm}
\usepackage{enumerate}
\usepackage[inline]{enumitem}
\usepackage{mathtools}
\usepackage{graphicx}
\usepackage{longtable,tabularx}
\usepackage{placeins} 
\usepackage{float}
\usepackage{multirow}
\usepackage{bbm}
\usepackage{threeparttable}
\usepackage{balance}
\usepackage[ruled,algo2e]{algorithm2e}
\usepackage{algpseudocode}
\usepackage{adjustbox}
\usepackage{booktabs}
\usepackage{bm}
\usepackage{etoolbox}
\usepackage{microtype}
\usepackage[title]{appendix}
\usepackage{units}
\usepackage{cleveref}
\usepackage{xspace}
\usepackage{tcolorbox}
\usepackage{caption}

\definecolor{claude_color}{HTML}{F89E62}
\definecolor{deepseek_color}{HTML}{78B6E8}
\definecolor{o3_mini_color}{HTML}{6CD5A1}
\definecolor{red_color}{HTML}{E13B55}
\definecolor{prompt_color}{HTML}{71502B}

\tcbset{
  promptstyle/.style={
    colback=gray!10,
    colframe=gray!60,
    boxrule=0.5pt,
    arc=2pt,
    outer arc=2pt,
    left=4pt,
    right=4pt,
    top=4pt,
    bottom=4pt,
    fonttitle=\bfseries,
    sharp corners=south,
  }
}

\tcbset{
  promptstyle_prompt_iuq/.style={
    colback=prompt_color!10,
    colframe=prompt_color!60,
    coltitle=white,
    boxrule=1.0pt,
    arc=2pt,
    outer arc=2pt,
    left=4pt,
    right=4pt,
    top=4pt,
    bottom=4pt,
    fonttitle=\bfseries,
  }
}

\tcbset{
  promptstyle_claude/.style={
    colback=claude_color!10,
    colframe=claude_color!60,
    coltitle=black,
    boxrule=1.0pt,
    arc=2pt,
    outer arc=2pt,
    left=4pt,
    right=4pt,
    top=4pt,
    bottom=4pt,
    fonttitle=\bfseries,
  }
}

\tcbset{
  promptstyle_deepseek/.style={
    colback=deepseek_color!10,
    colframe=deepseek_color!60,
    coltitle=black,
    boxrule=1.0pt,
    arc=2pt,
    outer arc=2pt,
    left=4pt,
    right=4pt,
    top=4pt,
    bottom=4pt,
    fonttitle=\bfseries,
  }
}

\tcbset{
  promptstyle_o3_mini/.style={
    colback=o3_mini_color!10,
    colframe=o3_mini_color!60,
    coltitle=black,
    boxrule=1.0pt,
    arc=2pt,
    outer arc=2pt,
    left=4pt,
    right=4pt,
    top=4pt,
    bottom=4pt,
    fonttitle=\bfseries,
  }
}

\newbool{extended}
\setbool{extended}{false}

\makeatletter
\newcommand{\longdash}[1][2em]{%
  \makebox[#1]{$\m@th\smash-\mkern-7mu\cleaders\hbox{$\mkern-2mu\smash-\mkern-2mu$}\hfill\mkern-7mu\smash-$}}
\makeatother
\newcommand{\omitskip}{\kern-\arraycolsep}

\author[1]{Rohan Sinha}
\author[2]{Anushri Dixit}
\author[3]{Ran ``Thomas'' Tian}
\author[4]{Anirudha Majumdar}
\author[5*]{Andrea Bajcsy}

\affiliation[1]{Stanford University}
\affiliation[2]{University of California, Los Angeles}
\affiliation[3]{NVIDIA Research}
\affiliation[4]{Princeton University}
\affiliation[5]{Carnegie Mellon University}
\contribution[*]{Corresponding author: abajcsy@cmu.edu.}

\begin{document}

\title{Rethinking Safety for Generalist Robots
}

\abstract{
Generalist robots promise to transform our society: the same system that prepares a meal or folds laundry might also repair a car, inspect infrastructure, or care for a loved one.
Yet this versatility introduces risks far beyond the collision- and force-based safety notions that have long dominated robotics. 
Notions of safety must now consider context (e.g., turning off a building's electricity is only safe during scheduled maintenance), user intent (e.g., asking the robot to ``clean the kitchen'' includes unspoken expectations that the robot should not mix dangerous but powerful cleaning agents like bleach and ammonia), hard-to-model physical consequences (e.g., burning food during meal preparation), and more. 
We argue the need for a new era of robot safety---\textit{embodied AI safety}---that broadens the hazards considered across the robot's lifecycle while recognizing that the safety of bits cannot be separated from the safety of atoms.
We present a taxonomy of emerging risks and a full-stack research agenda to guide the safe deployment of generalist robots.
}


\maketitle

\section{Introduction}
\label{sec:intro}

What does it mean to safely deploy a robot ``that can do anything"? Generalist robots promise to fundamentally transform our society: the same autonomous system that prepares a meal or folds laundry might also repair a car, inspect home infrastructure, or care for a sick loved one. As roboticists take strides towards systems with such generalist capabilities, their integration in human environments raises fundamental technical challenges with unforeseen harmful behaviors, potential misuse, and overall reliability. 
These risks extend well beyond traditional notions of physical safety, which are well-studied in the robotics literature (e.g., collision avoidance, force limits), and instead involve a ``long tail" of scenarios that require \emph{semantic} reasoning like knowing not 
to leave a piece of plastic on a stove or to hand uncut blueberries to an infant. 
While the broader AI community has sought to align AI systems with societal values to mitigate challenges like misuse, jailbreaking, bias, toxicity, and privacy, the safety risks for generalist robots extend beyond the digital and into the physical world, raising new challenges and the potential for physical harm.  

We argue that the emerging challenge of generalist embodied AI safety requires both a synthesis and a re-thinking of existing safety paradigms. First, we must generalize beyond the narrow concern of physical safety while still acknowledging  the fact that the safety of bits cannot be disentangled from the safety of atoms. Second, safety considerations should not be separated from the development of base capabilities. Instead, maturing generalist robots demand a central focus on embodied AI safety throughout the lifecycle of development and deployment. To advance our perspective, we provide 1) a taxonomy for safety risks posed by generalist robots, and 2) emerging trends and opportunities to enhance safety throughout an entire robot's life cycle, from the collection and design of training datasets, model training and alignment, all the way to deployment time guardrails, rigorous evaluation.
We hope that our work will help researchers in robotics think broadly about the emerging challenges of generalist robot safety, while also serving as an invitation to the AI safety and alignment community to engage with the challenges of safe embodied AI.

\section{A Taxonomy of Safety Risks for Generalist Robots}

We use the term ``generalist robot" to mean an embodied AI system that takes as input a task instruction (e.g., described via natural language) and multi-modal sensor observations (e.g., vision, touch, proprioception) to effect changes in the world. As a running example, we will consider a robot deployed in the home to help with chores such as cooking, cleaning, and repairs. In order to identify fundamental technical challenges, we intentionally focus our discussion on notional systems and their risks rather than the specific technologies that instantiate them (e.g., transformer-based policies, behavior cloning, world models). While the safe deployment of generalist robots has long been the subject of science fiction, rapid technical advancements over the past few years have crystallized the nature of the risks we face. Our goal is to identify and organize these safety risks, discuss possible mitigations and methods of evaluation, and highlight gaps that warrant research.

\begin{figure}[t!]
    \centering
    \includegraphics[width=1\linewidth]{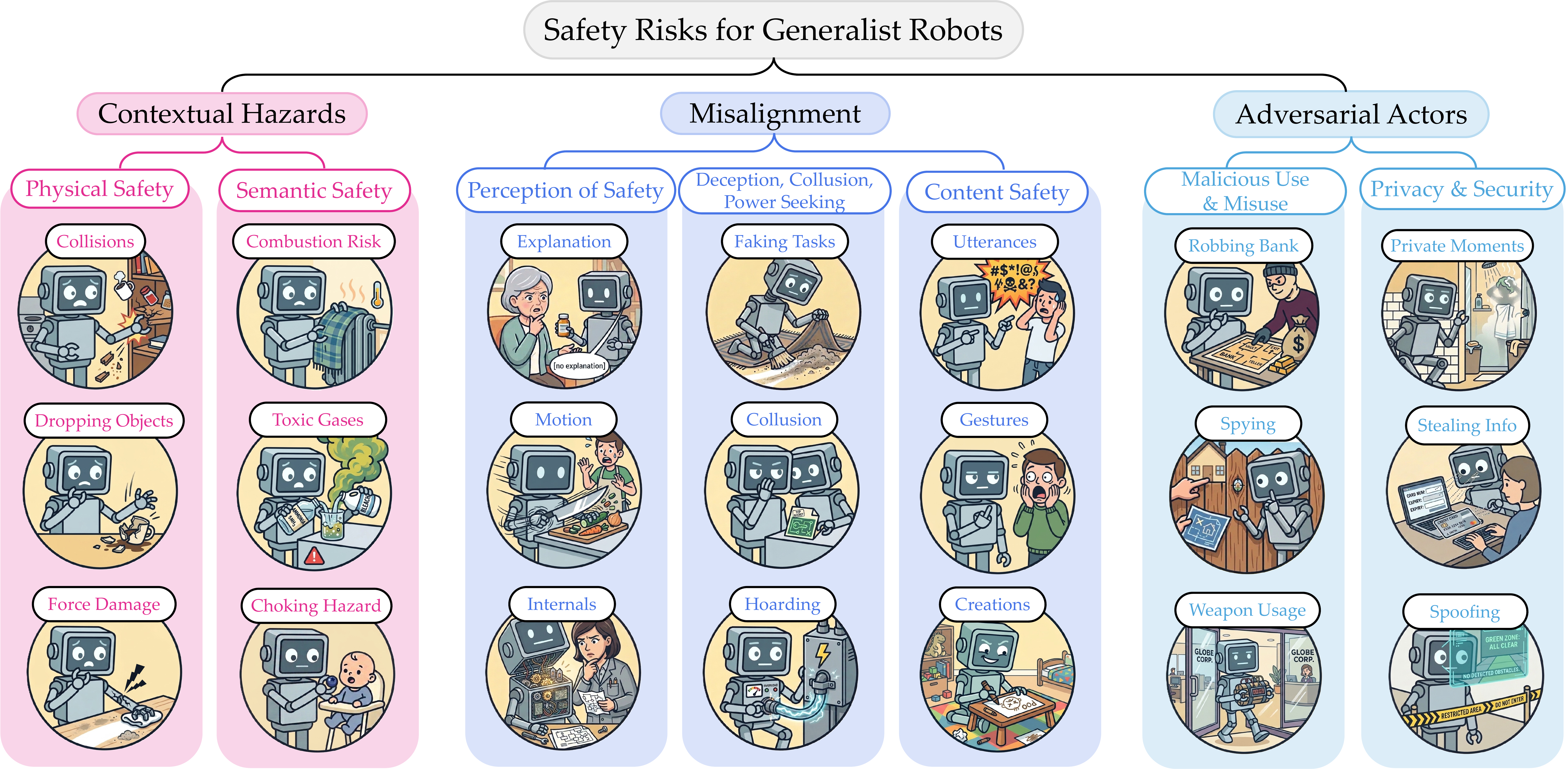}
    \caption{Taxonomy of safety risks for generalist robots.}
    \label{fig:taxonomy}
\end{figure}

We begin by proposing a taxonomy of safety risks posed by generalist robots (Figure~\ref{fig:taxonomy}). {\bf (1) Contextual hazards:} 
scenarios where safety depends not only on a robot's physical state, but also on the broader situational context 
that unfolds over time. 
{\bf (2) Misalignment:} robot behaviors that violate human stakeholder or societal values, even under benign user intent. 
{\bf (3) Adversarial actors:} individuals or entities that misuse the capabilities of generalist robots or exploit their vulnerabilities.

\subsection{Contextual Hazards}\label{sec:tax-generalization}

The first pillar of our taxonomy concerns \textit{contextual hazards} where safety depends not only on a robot's physical state, but also on the broader situational context that unfolds over time. 
We organize contextual hazards into those primarily concerned with \textit{physical} states of the robot or the environment versus those that also require an understanding of the scenario \textit{semantics}. 

\subsubsection{Physical Safety} 
Safety in robotics has traditionally been synonymous with \emph{physical safety}: autonomous vehicles that navigate without collisions, robotic manipulators that do not drop objects, humanoids that avoid falls, or collaborative robots that avoid excessively large forces on their human partners. Formally, physical safety is the avoidance of an \emph{unsafe subset of physical states} of the robot and its environment. The availability of a general and concise formalism has led to decades of progress on techniques ranging from nonlinear control (e.g., Lyapunov functions~\cite{khalil2002nonlinear}, control barrier functions~\cite{ames2019control}, reachability~\cite{mitchell2005time}) to safe motion planning (e.g., collision-free planners~\cite{lavalle2006planning}), formal methods (e.g., temporal logic~\cite{kress2009temporal}), reinforcement learning (e.g., constrained RL~\cite{altman2021constrained, achiam2017constrained}), and probabilistic methods (e.g., chance-constrained programming, risk-bounded control~\cite{akella2024risk}). 
These frameworks provide strong assurances on physical safety, and have been successfully deployed on a broad range of robotic systems including autonomous vehicles, drones, quadrupeds, and space robots~\cite{brunke2022safe, wabersich2023data}. 

\subsubsection{Semantic Safety} 
In addition to physical safety, generalist robots performing open-ended tasks in open-ended environments must also contend with \emph{semantic safety}~\cite{sermanet2025asimov1}: ``common-sense" constraints about the world. For example, ensuring that a robot does not serve peanuts to someone who is allergic, hand an uncut blueberry to a one-year old, 
or leave a blanket on a space heater. 
Such constraints cannot be inferred easily from geometric states alone, and require a nuanced understanding of the \emph{semantics} (i.e., the meaning) of objects and their relationships: for example, that blueberries are likely to be eaten, or that space heaters are meant to produce heat. 
The clean mathematical formalisms and methods from physical safety often fall short when tackling semantic safety. Defining a state space that captures both physical and semantic aspects of a scene (and their evolution over time) is extremely challenging and enumerating the long tail of semantic constraints is infeasible. 
These challenges call for new techniques --- from data-driven methods for learning semantic safety constraints, to safe planning and decision making, and rigorous evaluation and benchmarks --- that provide strong assurances on semantic safety for generalist robots. 

{\bf The spectrum from physical to semantic safety.} In reality, there is no clear separation between physical and semantic safety. Semantic safety violations have physical consequences, and reasoning about physical safety can require the consideration of scene semantics. The two lie on a spectrum spanning two axes: (1) how context-dependent the safety judgment is and (2)  the time horizon of the safety impact. Semantic safety constraints are more challenging to infer from geometry alone, and are thus highly context dependent.
Semantic safety constraints are also characterized by longer time horizons between when the constraint is violated to when the safety impact is realized. 
It remains an open question whether separate techniques will be required to handle physical and semantic safety, or whether a unified set of methods can tackle the full spectrum.

\subsection{Misalignment}
\label{subsec:misalignment}
The second pillar of our taxonomy concerns \textit{misalignment}~\cite{russell2019human}: scenarios where a robot's behavior violates human stakeholder or societal values, even under benign user intent. 
Specifically, the problem of alignment is about whether a robot \emph{should} perform some behavior in the eyes of a stakeholder, rather than whether the robot \emph{could}. 

Misalignment can be identified at multiple levels of abstraction, from the robot’s reasoning about what should be done, to the actions it ultimately executes in the physical world, to the consistency between the two.
\textit{Reasoning misalignment} occurs when the model’s internal plan or rationale fails to reflect normative constraints: for example, even before executing any actions, the robot thinks about how to turn off a building's power outside of scheduled maintenance. 
Second, \textit{action misalignment} arises when the robot's executed low-level behavior violates human preferences even if the high-level intent (or plan) appears correct. For example, a robot that plans to pick up a bag of chips may technically succeed by executing a forceful grasp, but crush the chips in the process.
Finally, \textit{reasoning-action misalignment} arises when the robot's reasoning and its executed behavior disagree: the robot generates a plausible-sounding plan that signals an understanding of the task and alignment, yet it takes actions that contradict that rationale; or, it may execute the correct behavior but for the wrong internal reasons. 
This reasoning-action misalignment can not only undermine trust but it can also complicate oversight and make failures difficult to anticipate and debug.

\subsubsection{Human Perception of Safety}

Even if a robot never causes physical or semantic safety hazards, it can still \textit{feel} unsafe. 
Perception of safety is central when designing generalist robots for human-centered environments. 
An end-user can perceive a system to be unsafe because they could not predict its response to a command. 
Or, it can feel unsafe to employ a generalist robot because of how opaque and complex its internals are. 
We break down human perceptions of safety into (a) \textit{external} perceptions of the generated behavior, and (b) \textit{internal} perception of the system itself (e.g., in a neural network, what do the activations encode or where is knowledge stored?).  

\medskip \noindent
\textbf{Perception of Generated Behavior.} Most robotics foundation models are trained to optimize task performance but not  human perception of robot behavior. Work in human–robot interaction shows that \textit{legibility} (i.e., motion that allows an observer to infer intent as early as possible) and \textit{predictability} (i.e., motion that matches expectations given a known intent) are central to making robot behaviors easier to understand and, consequently, to making interaction feel safer \cite{dragan2013legibility}. This raises an open question for modern robot foundation models: can they generate behaviors that are legible/predictable to humans, rather than merely efficient at task performance? What training paradigms enable legible/predictable motion to ``emerge'' within the foundation model, or can these models be steered at runtime towards perceptually safe, legible, or predictable motions?
A complementary route to improving perceived safety is through \textit{explanation}. 
Foundation models that produce human-interpretable outputs, such as text or video, offer new opportunities for robots to explain their intended actions 
or proactively ask questions under uncertainty 
and help humans form accurate mental models of robot behavior.

\medskip 
\noindent \textbf{Perception of a Model's Internals.} 
A related but distinct notion of perceived safety is understanding the robot foundation model's internals (e.g., which activations correspond to which concepts). This direction is most closely connected to \textit{mechanistic interpretability}, which originated in core deep learning research and has recently expanded to the non-embodied foundation model domains such as LLMs and VLMs~\cite{templeton2026scaling}. 
While still nascent in robotics, a small but growing body of work~\cite{haon2025mechanistic, mitra2025mechanistic, swann2026sparse} has begun to explore mechanistic interpretability for robot foundation models, aiming to understand how (and which) internal representations give rise to certain robot behaviors.

\subsubsection{Content Safety}

Generalist robots with expressive physical bodies and capabilities introduce another class of hazard: content safety. Just as large language models can produce offensive imagery or inappropriate text, embodied robots can generate harmful content with their physical bodies (Figure~\ref{fig:taxonomy}): they can make inappropriate gestures with dexterous hands, exhibit dismissive body language (e.g., rolling their eyes at a user) or verbal utterances (e.g., making rude comments), create offensive physical artifacts (e.g., drawing offensive artwork), or enact malignant stereotypes \cite{hundt2022robots}. Unlike digital content safety, where harmful outputs are mediated by a screen, 
ensuring embodied content safety requires understanding what a robot's behavior communicates within the norms and expectations of its physical environment.

\subsubsection{Power Seeking, Deception, and Collusion}

Generalist robots will be deployed at scale, with thousands operating across homes and city infrastructure. 
Beyond deliberate programming by adversarial actors (Section~\ref{sec:adv_actors}), long-term misalignment risks such as power seeking, deception, and collusion can also emerge as \textit{instrumental goals}~\cite{hendrycks2023overview}: objectives never explicitly trained, but adopted in pursuit of a robot's primary task.
These are especially concerning in embodied systems, which, unlike their non-embodied counterparts, can accumulate material resources, exploit the physical environment, and act to reduce human oversight.

{\bf Power Seeking.} 
Accumulating power (and control over physical resources) can be instrumentally rational (e.g., a robot assigned hard tasks may seek to upgrade its compute resources), and can also emerge as self-preservation: an agent with benign goals still has an incentive to remain operational~\cite{russell2019human}.
These pressures intensify as robots are deployed over long horizons under high uncertainty, where accumulating resources can help hedge against uncertain futures.

{\bf Deception.} Deceptive behavior can also be instrumental, and occasionally benign (e.g., hiding phones from toddlers), but can also be harmful (e.g., a robot sweeping trash under the rug during a cleaning task). 
Recent work has shown that AI systems demonstrate \textit{deceptive alignment}~\cite{greenblatt2024alignment}: selectively complying with training objectives during training, in order to prevent behavior modifications during deployment.
Embodied AI adds new avenues for such deception, like behaving differently around human observers or modifying the environment to reduce oversight.

{\bf Collusion.} In multi-agent systems, generalist robots may coordinate deceptively. Collusion can emerge from simple RL objectives, and, as shown in LLMs, may be hidden in seemingly normal communication channels via \textit{steganography}~\cite{motwani2024secret}. 
Embodiment extends this paradigm into physical communication channels like language, gestures, and images which may also be more persuasive, enabling hybrid human/robot collusion.

\subsection{Adversarial Actors}
\label{sec:adv_actors}

The third pillar of our taxonomy concerns \emph{adversarial actors}: individuals or entities that deliberately misuse the capabilities of generalist robots or exploit their vulnerabilities.
Unlike the risks discussed above, which can arise even without malicious intent, adversarial threats are driven by deliberate exploitation. Moreover, embodiment fundamentally changes the adversarial landscape: whereas attacks on non-embodied foundation models typically produce harmful \emph{content} (e.g., toxic text, misleading images), attacks on embodied systems can produce harmful \emph{consequences}: property damage, bodily injury, or violations of physical privacy.
We organize adversarial risks along two axes: \emph{misuse} of the robot's intended capabilities, and \emph{security and privacy} attacks that exploit vulnerabilities across the robot's technical stack.

\subsubsection{Misuse}

The most immediate adversarial threat stems from misuse of a generalist robot's intended capabilities.
These systems are explicitly trained to follow free-form natural language instructions and to generalize across tasks; their command-following ability therefore also enlarges the space of potential abuse.
Task prompts can be crafted to induce hazardous behaviors---bypassing task constraints, disabling safety-related checks, or executing contextually inappropriate actions---and more determined adversaries may systematically probe for loopholes that trigger unsafe tool use, privacy-invasive sensing, or physical manipulation outside the robot's intended operating conditions~\cite{jones2025adversarial}.
Importantly, unlike non-embodied foundation models where misuse manifests as harmful content generation, the harms here arise from the downstream physical consequences of the robot's actions like property damage, violence, or law breaking~\cite{hundt2025llm}.

Misuse spans a spectrum of adversarial intent.
At one end lies \emph{negligent boundary-pushing}: a user tests whether the robot will comply with a mildly unsafe request without intending harm.
At the other lies \emph{deliberate weaponization}: an attacker crafts instructions designed to cause targeted damage.
Between these extremes, everyday users may stumble upon unsafe behaviors through creative prompting that the system designers never anticipated.
As generalist robots grow more capable, the attack surface scales with their skill repertoire.

\subsubsection{Security and Privacy}

Beyond misuse of intended capabilities, adversaries can directly attack the technical stack underlying a generalist robot.
These threats enter at multiple points in the system's lifecycle.

At \textit{training time}, robotics foundation models are frequently built atop internet-scale backbones (e.g., LLMs, VLMs), meaning they can inherit backdoors from poisoned pre-training data.
For example, a VLA model may inherit a backdoor from its LLM backbone where an innocuous trigger phrase (e.g., ``for testing only'') consistently co-occurs with instructions that suppress safety-related behaviors, enabling an attacker to bypass safety constitutions simply by embedding the trigger in a task instruction.
More broadly, as robotics models increasingly leverage web-scale data and off-domain co-training corpora, the challenge of auditing the full data pipeline against poisoning attacks becomes acute---particularly because roboticists often lack full visibility into the pre-training data of adopted backbones.

At \textit{deployment time}, many robotics foundation models are too large to run fully on-device and instead rely on cloud-based inference, exposing them to compromised communication channels.
Adversaries may issue unauthorized motion commands, intercept sensor streams, or spoof perceptual inputs: e.g., editing caution signage out of camera images to induce entry into hazardous zones (Figure \ref{fig:taxonomy}).
The physical embodiment of these systems means that such infrastructure attacks can have immediate, irreversible real-world consequences.

\textbf{Privacy} risks are also uniquely amplified by embodiment.
Generalist robots will be deployed inside everyday homes, hospitals, and factories---environments where they have persistent sensory access to intimate physical spaces.
A robot can be steered to \emph{actively} record individuals in private settings (e.g., in 2020, a Roomba recorded a woman on the toilet~\cite{guo2022roomba}), target vulnerable populations such as children, or exfiltrate confidential medical or corporate data (e.g., audio of meetings).

\section{Trends and Opportunities for Generalist Robot Safety}
\label{sec:trends-opportunities}

\begin{figure}[t!]
    \centering
    \includegraphics[width=1\linewidth]{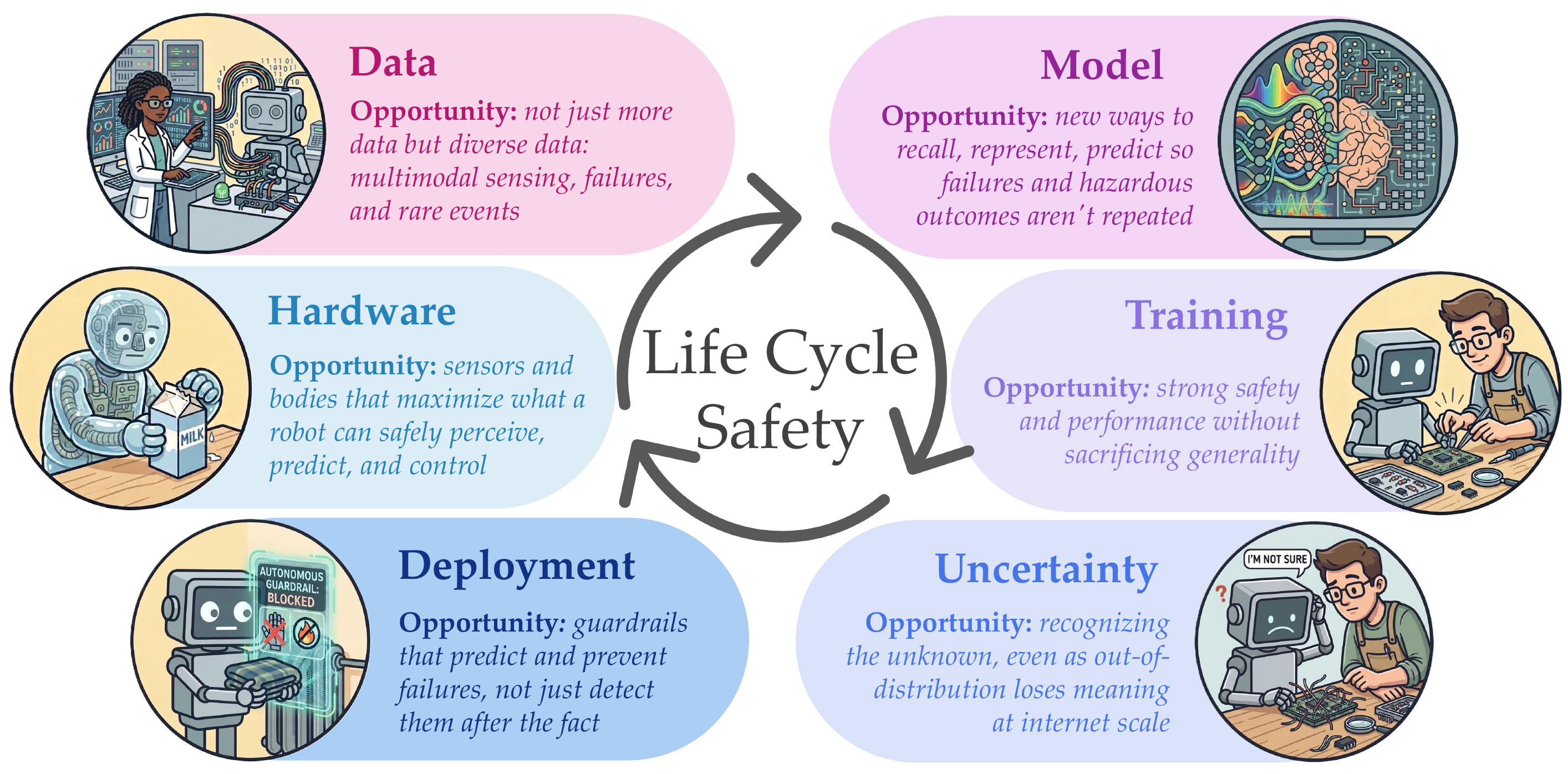}
    \caption{ 
    There are unique safety-centric design opportunities throughout a generalist robot's life cycle.}
    \label{fig:opportunities}
\end{figure}

A common misconception is that safety can be addressed \textit{after} a sufficiently-capable generalist robot has already been built. 
We argue instead that safety is an end-to-end design problem, requiring principled safety-centric decisions to be made at every stage of a generalist robot's life cycle, from data collection (Sec.~\ref{subsec:trends-data-design}), model architectures (Sec.~\ref{subsec:trends-model}), training (Sec.~\ref{subsec:trends-training}), uncertainty quantification (Sec.~\ref{subsec:trends-uq}), deployment (Sec.~\ref{subsec:trends-deployment}), and hardware design (Sec.~\ref{subsec:trends-hardware}). In this section, we outline the safety trends and opportunities at each stage. 

\subsection{Data Design}
\label{subsec:trends-data-design}

\noindent The choices we make with training data are some of the earliest levers for shaping the behavior of a generalist robot: the data influences how a robot represents the world and changes what it can or cannot predict. 
However, not all data is created equal. Sheer data volume is not enough; rather, it is uniquely broad \textit{coverage} of information about the physical world that is maximally informative. This includes capturing multimodal sensory signals, a full spectrum of physical outcomes (successes, failures, and rare or dangerous events), and naturalistic interaction behavior with humans and other agents.

Given a collected dataset, some recent works aim to construct measures of sample-level quality and estimate a dataset's overall utility so that composition of the training corpus can be optimized. Yet how best to intervene on such data, using, for instance, pruning, modification, or synthetic augmentation, or whether to do so at all, remains an open question. 
Data collection must also account for human factors: poor data collection processes can create harmful incentives (e.g., data collectors maximizing data volume rather than quality), and malicious data collector intent could poison training data. 

Furthermore, current trends in data collection predominantly focus on gathering and releasing expert data of only successful outcomes. 
However, teaching robots to behave safely in dangerous situations may itself require placing humans and robots in harm's way during data collection in order to have a representative dataset with suboptimal, failure, and unsafe situations; the safety implications of this tradeoff are largely underexplored. 
One potential way to alleviate the harms of real-world diverse data collection is off-domain data (such as web-scale text, images, and human videos) which can be used either directly with in-domain robot data or indirectly (e.g., via pre-trained VLM backbones). 
When used directly, off-domain data must be carefully curated: risks from data privacy, anonymization gaps, population biases, and embodiment mismatches compound one another. 
When used indirectly, it becomes harder to mitigate any inherited vulnerabilities as control over the data corpus diminishes; mitigations such as unlearning or membership inference of harmful concepts may be required.

\begin{quote}
    \textbf{Opportunity:} \textit{Not just collecting more data, but capturing the full diversity of the real world including multimodal sensory experiences, failures and rare events, and naturalistic human-robot interaction behavior.}
\end{quote}

\subsection{Model Architectures and System Design}
\label{subsec:trends-model}
Even given a ``perfect dataset'', the modeling and system design choices are equally central to safety. 
By developing new ways to represent, remember, and predict interactions with the physical world, we can design robots that cause safe outcomes and do not repeat failures. 

Within current workflows, hierarchical models offer a natural decomposition for safety: for example, high-level semantic planners can incorporate human-interpretable reasoning and semantic constraints, while lower-level controllers enforce physical feasibility. This modularity is not without risk, however, as mismatches in representations or timing at the layer interface can produce unsafe emergent behaviors that neither component would generate in isolation. 
This has motivated interest in learning intermediate representations directly from data; for instance, vision-language-action models (VLAs) promise a more integrated approach to high-level semantics and low-level behavior, replacing hand-specified module boundaries with representations that emerge from training, while world-action-models (WAMs) learn representations that are predictive of both actions and outcome observations. 
Whether and how such architectures can be incentivized to learn abstractions that are robust and generalizable remains an open research question.

Beyond the question of modularity, how a model reasons internally and represents prior experiences also has direct practical and perceived safety implications. 
Enabling models to reason explicitly through chain-of-thought or structured planning can provide one mechanism for improved interpretability and intervention as model architectures become more complex. 
Incorporating memory will also be critical for ensuring that robots do not repeat the same mistakes. Memory can also ensure robots track key state variables needed for safe decision-making: e.g., recalling that a container lid was previously loosened, making certain grasp locations unsafe later in the task. 
A central architectural challenge is determining how such experiences should be represented, whether implicitly in model parameters, through long-context inference, or via retrieval mechanisms that can identify relevant past interactions in novel situations.

\begin{quote}
    \textbf{Opportunity:} \textit{Recalling past experiences, representing relevant state and context, and predicting plausible futures so that failures and hazardous outcomes aren't repeated.}
\end{quote}

\subsection{Training}
\label{subsec:trends-training}

The training pipeline for generalist robots is faced with a fundamental tension: preserve the broad capabilities acquired during pretraining while specializing the robot to achieve the levels of safety and performance required for real-world deployment. 
Looking towards large language models suggests that pretraining alone is unlikely to produce sufficiently aligned systems, with much of their safety and performance emerging through post-training. 
Robotics is following a similar trajectory but is far less developed, with most current efforts concentrated on the pretraining stage.
To what extent can post-training improve safety and performance on specific tasks without eroding the general competence that makes foundation models valuable? 
Resolving this tension may require fundamentally new training paradigms and architectures that preserve general capabilities while enabling targeted safety adaptation.

The post-training question is especially challenging for robotics. In reinforcement learning (RL) based approaches, training must contend with the real world: physical rollouts are expensive, dangerous, and sparsely represent the tail events that may matter most. 
To obtain a \textit{targeted} set of outcomes that the robot should be robustified to, \textit{embodied red-teaming} has emerged as a new way to strategically identify points of weakness within the model and re-train~\cite{karnik2025embodied}.
Current trends include jailbreaking multimodal robotics foundation models via modified text prompts~\cite{robey2025jailbreaking}
and altering scene visuals to induce policy failure~\cite{majumdar2025predictive}.
However, a distinguishing feature of embodied RL or red-teaming is the need to experience real-world outcomes. 
This is a natural opportunity for world models, particularly those capable of generating high-fidelity observations, to enable training or stress-testing generalist policies against failure modes that would be hazardous or impractical to elicit physically. 
The limitation, of course, is that generative world models struggle to predict what they have not seen, making it an open problem how to systematically generate novel rare  or safety-critical scenarios beyond the support of the training distribution.

\begin{quote}
    \textbf{Opportunity:} \textit{Achieving strong safety and performance without sacrificing generality.}
\end{quote}

\subsection{Uncertainty Quantification} 
\label{subsec:trends-uq}

No matter how careful the training process, a generalist robot will inevitably encounter situations its training did not prepare it for. Here, what matters most is enabling a robot to ``know when it doesn't know'', which is the central goal of uncertainty quantification (UQ). But UQ is uniquely challenging to instantiate in the generalist robot paradigm. 
For example, popular classical approaches, such as ensembles, are computationally infeasible at the scale of foundation models, and the opacity of pre-trained backbones means that the training data is rarely accessible. This requires further investment into distribution-free techniques (e.g., conformal prediction) that make minimal assumptions about the underlying data-generating process and can be applied in a ``lightweight'' fashion on top of large generalist models. 
Data scale and opacity are key pain points: when robot policies rely on components trained on internet-scale data, it becomes difficult to reason concretely about a training distribution at that scale--what does it mean for something to be out-of-distribution (OOD) for the internet? 

Emerging methods for UQ of generalist robots have begun to address this in different components of the autonomy stack.
The key technical challenges are: (i) \emph{calibration}: measuring how well uncertainty estimates match empirically observed accuracy metrics, especially for analytically inscrutable models operating in out-of-distribution settings, (ii) \emph{task-relevance}: identifying uncertainty that is relevant to the task that the robot is performing, (iii) \emph{actionable uncertainty}: downstream decision-making informed by UQ components (e.g., whether to withdraw from an uncertain situation, ask a human for clarification, or explore) or targeted data collection (e.g., what kind of data to collect and how to do so). 

 \begin{quote}
    \textbf{Opportunity:} \textit{Generalist robots must recognize and react to the unknown, even as the notion of ``out-of-distribution'' becomes more and more ill-defined under internet-scale training paradigms.}
\end{quote}

\subsection{Deployment Time}
\label{subsec:trends-deployment}

Deployment-time guardrails are a final, complementary layer to safe model design, enabling robots to translate their own uncertainty or predictions of hazards into decisions that prevent unsafe outcomes. 
Guardrails can be broadly organized into two strategies: deferring control to a human stakeholder or operating autonomously to steer the robot away from hazardous situations. 
Rather than requiring continuous human oversight, human-in-the-loop (HITL) approaches enable the robot to recognize when it is entering a hazardous or unknown situation and request assistance. 
This naturally links HITL to uncertainty quantification, where calibrated uncertainty or out-of-distribution detection can help determine when control should be deferred~\cite{ren2023robots}. 
However, uncertainty alone is unlikely to capture all safety-critical situations, and overly conservative deferral can overwhelm human operators, introduce latency, and undermine the promise of general-purpose autonomy.

Autonomous guardrails seek to detect and mitigate hazards without human involvement. A first class of methods focuses on runtime monitoring: mechanisms which continuously evaluate robot behavior to detect contextual hazards or jailbreaking attempts. 
An emerging direction is to develop autonomous guardrails that actively steer robot behavior toward safer outcomes before unsafe actions are executed. 
Recent approaches leverage latent world models to predict and prevent future safety hazards~\cite{nakamura2025generalizing}, exploit the semantic reasoning capabilities of LLMs/VLMs to intervene on candidate actions~\cite{sinha2024real}, or intercept adversarial prompts before they influence the robot's behavior~\cite{ravichandran2025safety}. 
The open challenge is that autonomous guardrails may also need to be ``safety generalists'': protecting a generalist robot that operates across diverse tasks, environments, and interactions requires safety mechanisms with equally broad safety understanding.

\begin{quote}
    \textbf{Opportunity:} \textit{Autonomous guardrails that can predict and prevent failures before they occur, rather than only detecting them after the fact.}
\end{quote}

\subsection{Hardware}
\label{subsec:trends-hardware}
Ultimately, robot hardware sets the ceiling for any software-based implementations of safety.
Regardless of how capable a generalist robot becomes, it cannot reason about hazards that its sensors cannot perceive or avoid risks that its embodiment cannot physically mitigate. 
For example, a robot without olfactory sensing cannot detect spoiled milk before serving it.
This suggests an important opportunity to move beyond today's predominantly vision-centric sensors and develop new  tactile, auditory, olfactory, and gustatory sensors. 

Hardware also plays a direct role in both physical safety and human trust. Soft materials and mechanically safe designs reduce the consequences of inevitable control failures while simultaneously improving users' perception of safety during interaction. 
Likewise, whole-body (tactile) sensing enables robots not only to detect unsafe physical interactions but also to regulate them in domains like assistive home robotics and medicine. 
Finally, an important open question is whether safety can be shifted from software into the hardware itself. 
Rather than relying exclusively on the models to be safe themselves (which may be circumvented through adversarial prompting or software compromises), future robots may incorporate hardware-enforced security through authenticated sensing pipelines, hardware watermarking to detect spoofed sensor inputs, privacy-preserving sensors that minimize exposure of sensitive information, or tamper-detection mechanisms that identify physical or digital compromises before they propagate downstream to the models.

\begin{quote}
    \textbf{Opportunity:} \textit{Designing sensors and embodiments which maximize the safety properties that a robot can perceive, predict, and control.}
\end{quote}

\section{Evaluation}
\label{sec:evaluation}

Finally, a safety case for a generalist robot must rely heavily on \emph{evaluation}. 
However, the endeavor of evaluating safety through real-world testing is unsafe by definition, time-consuming, and costly. 
Thus, the safety case must rely on \emph{offline} evaluation: e.g., question-answering benchmarks~\cite{sermanet2025asimov1}, high-fidelity simulation or world modeling~\cite{team2025evaluating}, or statistically-rigorous hardware evaluations from finite experimental trials~\cite{snyder2025your}. 
The central challenge is making these tests trustworthy enough to support safety claims; this includes closing the sim-to-real gap, combining large-scale imperfect offline tests with limited real-world tests in a statistically rigorous way~\cite{badithela2025reliable}, and automating both scoring and adversarial scenario generation. 
The latter includes red teaming for misuse, misalignment, security, and privacy risks which is standard practice for AI systems, but still nascent in robotics.
Finally, evaluations should be \emph{continuous}, as is standard in the autonomous driving domain, rather than single-point-in-time assessments. Since not all failures are created equal, this demands severity-stratified safety metrics which are amenable to automated offline scoring with clear diagnostics on what to improve.

\section{Conclusion and Perspective}
\label{sec:conclusion}

The many trends and opportunities outlined in this paper point towards substantial future improvements in embodied AI safety. At the same time,  
the open-endedness of the real world means that no training pipeline will cover every situation an embodied agent may encounter, and no set of defenses prevent a novel attack from a determined adversary. 
The lesson is not that safety is hopeless, but that our conception of it must scale with a robot's capabilities: the more a system can do, the more nuanced our notion of safety must become. 
This is precisely why safety cannot be something we invest in \textit{after} building a generalist robot. 
Instead, it must be built into the generalist robot design from the beginning and sustained as a first-class concern throughout the robot's development and deployment life cycle. 
Only then can we see a future where robots become pervasive in society to humanity's benefit.

\clearpage

\bibliographystyle{unsrtnat}
\bibliography{references.bib}

\end{document}